\documentstyle[12pt]{article}
\newlength{\extraspace}
\newlength{\extraspaces}
\newcommand{\be}{\begin{equation}
\addtolength{\abovedisplayskip}{\extraspaces}
\addtolength{\belowdisplayskip}{\extraspaces}
\addtolength{\abovedisplayshortskip}{\extraspace}
\addtolength{\belowdisplayshortskip}{\extraspace}}
\newcommand{\ee}{\end{equation}}

\newcommand{\ba}{\begin{eqnarray}
\addtolength{\abovedisplayskip}{\extraspaces}
\addtolength{\belowdisplayskip}{\extraspaces}
\addtolength{\abovedisplayshortskip}{\extraspace}
\addtolength{\belowdisplayshortskip}{\extraspace}}
\newcommand{\ea}{\end{eqnarray}}

\newcommand{\newsection}[1]{
\vspace{15mm}
\pagebreak[3]
\addtocounter{section}{1}
\setcounter{equation}{0}
\setcounter{subsection}{0}
\setcounter{footnote}{0}
\begin{center}
{\Large \thesection. #1}
\end{center}
\nopagebreak
\medskip
\nopagebreak}

\newcommand{\newsubsection}[1]{
\vspace{1cm}
\pagebreak[3]

\addtocounter{subsection}{1}
\noindent{ \sc \thesubsection. #1}
\nopagebreak
\vspace{2mm}
\nopagebreak}
\newcommand{\ie}{{\it i.e.}}

\newcommand{\is}{\! & \! = \! & \!}
\newcommand{\nonu}{\nonumber \\[.5mm]}

\newcommand{\half}{\textstyle{1\over 2}}
\newcommand{\delbar}{\overline{\partial}}

\newcommand{\zbar}{{\overline{z}}}

\newcommand{\R}{{R}}

\newcommand{\tpi}{{2 \pi i}}
\newcommand{\tV}{\widetilde{V}}
\newcommand{\tO}{\widetilde{O}}
\newcommand{\tJ}{\widetilde{J}}
\newcommand{\tP}{\widetilde{\phi}}
\newcommand{\Ph}{{\mbox{\raisebox{.1ex}{$\phi$}}}}
\newcommand{\rg}{{\mbox{\raisebox{.1ex}{$g$}}}}
\newcommand{\tg}{\widetilde{\rg}}
\newcommand{\ttg}{\tilde{g}}
\newcommand{\tW}{\widetilde{W}}
\newcommand{\gl}{n_{gh}}
\newcommand{\cbar}{{\overline{c}}}
\newcommand{\bbar}{{\overline{b}}}
\newcommand{\mubar}{{\overline{\mu}}}
\newcommand{\Lbar}{{\overline{L}}}
\begin{document}
\addtolength{\baselineskip}{.7mm}

\thispagestyle{empty}

\begin{flushright}
{\sc IASSNS-HEP}-92/5\\
February 1992
\end{flushright}
\vspace{.6cm}

\begin{center}
{\large{\sc{THE MASTER EQUATION OF}\\[3mm]
           {2D STRING THEORY}}}\\[1.5cm]

 {\sc Erik Verlinde}
\footnote{supported by the W.M. Keck Foundation.
          $\quad$ bitnet address: verlinde@iassns}\\[3mm]
{\it School of Natural Sciences\\[2mm]
Institute for Advanced Study\\[2mm]
Princeton, NJ 08540} \\[2cm]

{\sc Abstract}\\[1cm]

{\parbox{13cm}{A general method is presented for deriving on-shell
Ward-identities in (2D) string theory. It is shown that all
tree-level Ward identities can be summarized in a single
quadratic differential equation for the generating function
of all amplitudes. This result is extended to loop amplitudes
and leads to a master equation {\it \`{a} la} Batalin-Vilkovisky
for the complete partition function.}}
\end{center}

\noindent

\vfill

\newpage

\setcounter{section}{0}
\setcounter{equation}{0}

\newsection{Introduction and summary}

A important feature of 2D string theory is that
in addition to the usual physical operators at ghost level $\gl=2$
there are other BRST-invariant operators at different ghost levels
\cite{LZ,SM,PB}. The importance of this fact was emphasized by
Witten, who showed that the $\gl\!=\!0$ states constitute a
ring, called the ground ring, and that the operators
with $\gl\!=\!1$ are associated with symmetries \cite{EW}.
These symmetries were discovered previously
in the context of matrix models \cite{matrix},
and were independently
considered by Klebanov and Polyakov \cite{KP}

\newcommand{\bm}{b_0^-}

In the current literature most attention is given to
the states at ghost level $0$, $1$ and $2$, but there are
equally many physical states at other ghostlevels.
In fact, for every physical state $\Ph_I$ with $\gl \!=\!n_I$
there is an associated `anti-state' at $\gl=5-n_I$ given by
\be
\label{antistate}
|\tP^I\rangle=\bm|\Ph^I_c\rangle
\ee
where $\bm=(b_0-\bbar_0)$ and $\Ph^I_c$ is the BRST-invariant state
that is conjugate to $\Ph_I$.
In this paper will use these anti-states
to give a systematic derivation of the Ward-identities and
discuss the generalization to loop amplitudes. We find
that the information about all Ward-identities (and more) can
be summarized in a single equation for the generating
function $F=-\lambda^2\log Z$ of all string amplitudes.
It takes the form
\be
\label{meqn}
-\lambda^2
\sum_I {\partial^2 F\over\partial \rg^I\partial \tg_I}
+\sum_I {\partial F\over\partial \rg^I}{\partial F\over\partial \tg_I}=0
\ee
where $g^I$ and $\tg_I$ are the couplings for the states and
anti-states respectively and $\lambda$ is the string coupling constant.
This equation is known in the Batalin-Vilkovisky formalism as
the  the (quantum) master equation \cite{BV}.

In section 2 we review the spectrum of physical states
of 2D string theory and discuss the role of the corresponding
charges in relation with the symmetries and the perturbations
of the BRST-charge. The anti-states are introduced in section
3. We find that the structure constants of the symmetry
algebra of all charges are given by the $n$-point amplitudes.
The Ward-identities are derived  in section 4. Finally, in section
5 we write these identities in the form of the master equation,
and we discuss some aspects of the Batalin-Vilkovisky formalism.

\newsection{Physical states and (un-)broken symmetries}

In aim of this paper is to derive Ward identities associated
with (unbroken) symmetries of string theory.
Our motivation comes from the recent studies of 2D string theory,
which showed that 2D strings have many unbroken symmetries.
The structure of the physical spectrum and symmetries
of 2D string theory in a flat background
has been significantly clarified in the recent work of
Witten and Zwiebach \cite{WZ}.
In the first two subsections we will summarize some of their
results, in particular the
construction of the symmetry charges\footnote{I thank E. Witten and
 B. Zwiebach for explaining these results to me before finishing
their manuscript.}. In section 2.3 we use this to study
the perturbations of the BRST-charge.

\newsubsection{The spectrum of 2D string theory}

First we summarize the spectrum for the
uncompactified string and we
restrict our attention to the positive branch of
Liouville momenta $p_L\geq Q_L$ ($Q_L$ is the
background charge). It is known for a long time that
the spectrum of physical states contains an infinite set of
discrete states at special values of the momenta \cite{discrete},
These discrete states, which also appear in matrix models \cite{dismat},
have been suggested to have a topological origin \cite{Polyakov}.
More recently it was discovered that
each of the discrete states is accompanied by a state $O$
at ghost level $\gl=0$ and a pair of states $J^L$ and $J^R$
at $\gl=1$ \cite{LZ}-\cite{EW}. Together with the tachyon
vertex operators $V_p$,  this gives the following set of
physical states
\ba
\label{spec}
\gl=0 & : & O_{lm}\nonu
\gl=1 & : & J_{lm}\\ \nonumber
\gl=2 & : & V_{lm}, V_p,
\ea
where $p\in \R$, and
$l$ and $m$ are the usual $SU(2)$ isospin labels.
These states are obtained by simply taking the
product of the left- and right-moving BRST-invariant states.
For example for $l=0$, we have
\be
\label{example}
O_{0,0}=1,\quad J^L_{0,0}=c\partial X,\quad
J^R_{0,0}=\cbar\delbar X,\quad V_{0,0}=c\cbar\partial X\delbar X,
\ee
where $X$ is the string coordinate. In fact, (\ref{spec})
only represents half of the physical states for $p_L\geq Q_L$.
By using the Liouville field $\varphi_L$ one can construct
new states by \cite{WZ}\footnote{Here and in the following
$[\cdot,\cdot]_{\pm}$ is an anti-commutator when both operators have odd
ghost-number and a commutator otherwise}
\be
\label{Qphi}
W=[Q,\varphi_L V]_{\pm}.
\ee
These states are clearly BRST-invariant, but not BRST-exact
since $\varphi_L$ itself is not a good operator.
We obtain in this way physical
states $W_p$ and $W_{lm}$ at ghost level $3$, for example
\be
W_{0,0}=(c\partial c \cbar+c\cbar\delbar \cbar)\partial X\delbar X.
\ee
In the same way one can construct new states at ghost level 1 and 2
from $O_{lm}$ and $J^{L,R}_{lm}$. The precise details of this construction
will not be important for our discussion.

A similar list of operators can be made for the compactified string models:
the tachyons carry discrete momenta $(p_L,p_R)$ and there are additional
discrete states with quantum numbers.
While the tachyon spectrum varies continuously with $R$, one finds that
the spectrum of discrete states behaves in a rather chaotic way.
Only  at the $SU(2)$-point the discrete states are labelled by
the $SU(2)_L\times SU(2)_R$ quantum numbers $(l,m_L, m_R)$.
When we perturb the radius $R$ away from its self-dual value $R_{SD}$,
all discrete states, except those at $m_L=m_R=0$ disappear,
but at rational values of $R/R_{sd}$ some of the
discrete states at non-zero $m_L$ and $m_R$ will re-appear.
One of the aims of this section is to obtain a better understanding of
this phenomenon.

\newsubsection{Conserved charges and symmetries}

Our following discussion is valid for any background
of the 2D string theory, and so we will use a more general notation.
For all known backgrounds there are BRST-invariant states at $\gl=0,1,2,3$
which we denote by $O_\alpha,J_a, V_i$, and $W_A$, respectively.
In the subsequent sections we write them collectively as
\be
\label{fields}
\Ph_I  = O_\alpha,\,J_a,\, V_i,\,
W_A.
\ee
These operators commute when their ghost-number
is even, and anti-commute when it is odd. We note
that not all operators are factorizable in left and right movers, nor
do they always have the same left and right ghost number. In this and other
respects $2D$ string theory has similar features as topological
field theory, and it is convenient to borrow some of the formalism of the
latter. In particular, one can apply the descent equations to construct
from a dimension zero field $\Ph$ with $[Q,\Ph]_\pm=0$
an associated current (=1-forms) $\Ph^{(1)}$ and two-form operator
$\Ph^{(2)}$ through\cite{WZ}
\ba
d\Ph^{(0)} \is [Q,\Ph^{(1)}]_{\pm},\nonu
d\Ph^{(1)}\is[Q,\Ph^{(2)}]_{\pm}.
\label{descent}
\ea
Note that this defines $\Ph^{(1)}$ and $\Ph^{(2)}$ up to BRST-commutators.

There are several ways in which one can derive relations among the different
states. One possibility is to consider the ground ring formed by the zero
forms $O_\alpha$ of the ghost number $0$ \cite{EW,KMS}. A second approach,
which is the one we will study in this paper, is to use the charges
constructed from the one-forms $\Ph^{(1)}$.
In this section we will examine these charges
associated with $J_a$ and $V_i$. The contour-integrals
of $J_a^{(1)}$ describe BRST-invariant observables
with ghost-number zero
\be
\label{Qa}
Q_a ={1\over\tpi}\oint J_a^{(1)}.
\ee
It can be shown that these charges commute  with $\bm$, and so
they generate symmetries between physical states \cite{WZ}
\be
\label{symm}
Q_a |V_i\rangle =f_{ai}{}^j |V_j\rangle.
\ee
This equation holds at the level of BRST-cohomology classes,
so we ignore any BRST-trivial contributions.
The symmetry charges form a closed algebra
\be
[Q_a,Q_b] =f_{ab}{}^c Q_c.
\ee
For example for the flat uncompactified
background the symmetry algebra
is related to area-preserving diffeomorphisms
and takes the form  \cite{EW,KP}
\be
[Q_{lm},Q_{l^\prime m^\prime}]=
((l+1)m^\prime-(l^\prime+1) m)Q_{l+l^\prime, m+m^\prime}.
\ee
These symmetries
represent the unbroken gauge-symmetries of the 2D string,
as will become clear below.

\newsubsection{The perturbed BRST-charge.}

Similarly as for the currents $J_a$ we can also use the 1-forms of
the operators $V_i$ to construct charges
\be
\label{Qi}
Q_i=\oint V^{(1)}_i.
\ee
These anti-commuting charges have ghost number 1 and map the
usual physical operators $V_i$ on to the states at $\gl=3$
\be
\label{VVW}
Q_i|V_j\rangle=C_{ij}{}^A |W_A\rangle.
\ee
The coefficients $C_{ij}{}^A$ are directly related to the
quadratic $\beta$-function, which indicates that the charges
$Q_i$ play an important role in studying the perturbations of
the model. As usual the physical perturbations of the action
are represented the integrated two-forms $V_i^{(2)}$
\be
\label{pert}
S^\prime=S+\delta t^i\int V^{(2)}_i.
\ee
The perturbed BRST-charge is determined by a simple
application of the Noether procedure as follows. Let us perform
a BRST-transformation with a coordinate-dependent parameter $\epsilon$
in (\ref{pert}). Using the descent equation (\ref{descent}) we find
\be
\delta_{{}_{BRST}} S^\prime=\int
\!\epsilon\, d(j_{{}_{BRST}}+\delta t^i V_i^{(1)}).
\ee
{}From this we read off the perturbation of the BRST-current
$j_{{}_{BRST}}$, from which we construct
the perturbed BRST-charge. We find
\be
\label{pertQ}
Q^\prime=Q+\delta t^i Q_i.
\ee
For example, for a vertex operator of the usual form $V=c\cbar\Psi$,
with $\Psi$ a dimension $(1,1)$ (matter) primary field,
the variation of the BRST-charge is $$\oint (dz\cbar \Psi+d\zbar c\Psi).$$

What happens to the symmetries under these perturbations.
When the perturbation $\delta t^iV_i$ does not commute
with one or more of the charges $Q_a$ the
symmetry will be, at least partly, broken. As will be explained in the
following section, we have the relation
\be
\label{goldstone}
Q^\prime |J_a\rangle=\delta t^i Q_a|V_i\rangle.
\ee
The state on the r.h.s. can be identified with
the Goldstone mode. When it is non-vanishing
we conclude from (\ref{goldstone}) that after
the perturbation the symmetry current $J_a$ is no longer physical.
At the same time equation (\ref{goldstone}) tells that the Goldstone
mode becomes a longitudinal mode, (=BRST-exact). This is the true
signature of a spontaneously broken gauge-symmetry.
In fact, this is the phenomenon that is responsible for the
disappearance and re-appearance of the discrete states.

\newsection{Anti-states and more symmetries}

In this paper we will work with
vanishing cosmological constant, and so, to get non-vanishing
amplitudes, we have to include the states on the
negative branch of Liouville momenta. On this branch the
non-vanishing BRST-cohomology is at ghost-levels $\gl=5,4,3,2$.
 We denote them by
\be
\label{antifields}
\tP^I=\tO^\alpha,\tJ^a, \tV^i,\tW^A.
\ee
They are related to the
the states $\Ph_I$ given in (\ref{fields})  by
\be
\tP^I=\bm \Ph^I_c,
\ee
where $\langle\Ph_c^J|\Ph_I\rangle=\delta_I{}^J$.
We will refer to $\tP^I$ as  the `anti-state' or `anti-field'
corresponding to $\Ph_I$ because as will become clear they describe
the on-shell modes of the space-time anti-fields \cite{Thorn}.\footnote{
Strictly speaking the space-time anti-fields
correspond to states with $\gl \geq 3$, and therefore it may
be more appropriate to associate $W_A$ with an anti-field and
$\tW^A$ with a field. Fortunately this distinction will not be essential
for our purposes.} In the subsequent sections all fields $\Ph_I$
and anti-fields $\tP^I$ will be treated on equal footing.
\newcommand{\tf}{{\widetilde{f}}}

Again we use the descent equations (\ref{descent})
to construct BRST-invariant charges from the fields $\Ph_I$
\be
Q_I ={1\over\tpi}\oint \Ph_I^{(1)},
\ee
which except for $Q_a$ have ghost number different from zero.
This means that they not really generate physical symmetries, but they
do lead to relations among the amplitudes.
As can be seen from (\ref{descent}) the charges $Q_I$
are conserved only up to BRST commutators. This fact
will become important later on.
In a similar way one can construct charges $\widetilde{Q}^I$ for the
anti-states, but these will not be considered in this section.

Both the states as their anti-states represent `good' operators in the
sense that they are annihilated by $\bm$ \cite{PN}.
Due to this fact all (genus zero) two-point functions vanish, but
the  three-point functions
\ba
\label{3pt}
\langle \Ph_I\Ph_J\tP^K\rangle \is f_{IJ}{}^K   \nonu
\langle \Ph_I\tP^J\tP^K\rangle \is \tf_I{}^{JK}
\ea
are in general non-vanishing provided the total ghost-number
adds up to $6$. These three-point functions describe the
action of the charges $Q_I$ on states
\ba
\label{rep}
Q_I|\Ph_J\rangle \is f_{IJ}{}^K|\Ph_K\rangle,\nonu
Q_I|\tP^K\rangle \is f_{IJ}{}^K|\tP^J\rangle + \tf_I{}^{KJ}|\Ph_J\rangle,
\ea
as well as the (anti-)commutation relations among the charges
\be
[Q_I,Q_J]_{\pm}=f_{IJ}{}^K Q_K.
\ee
Both these relations are valid modulo BRST-commutators. They are derived
by some  straightforward manipulations involving the $b$-fields,
that at the same time  show the consistency with the
$\bm$-constraint\footnote{I thank R. Dijkgraaf for helpful discussions
on this point.}. Notice that the coefficients $f_{IJ}{}^K$ are
(anti-)symmetric in {\scriptsize $I$} and  {\scriptsize $J$}.
We already made use of that in (\ref{goldstone}).

Requiring that the states $\Ph_I$ form a
representation of the algebra gives the relations
\ba
\label{jacobi}
0\is f_{IJ}{}^M f_{MK}{}^L  + \ \mbox{cycl. in {\scriptsize{$I$, $J$, $K$}},}\\
\label{relat}
f_{IJ}{}^M \tf_{M}{}^{KL}  \is \tf_{I}{}^{KM} f_{MJ}{}^L+
\ \mbox{(anti-)symm. in {\scriptsize{$I$, $J$ $\&$  $K$, $L$}}.}
\ea
An alternative way of arriving at the same
relations is to consider Ward-identities.
For example the Jacobi identity (\ref{jacobi}) follows from
\newcommand{\toint}{\oint\!}
\be
\label{ward}
\Bigl\langle \int\!
d\Ph^{(1)}_I \,\Ph_J\Ph_K\tP^L\Bigr\rangle=0,
\ee
which leads to
\be
\Bigl\langle \Bigl(\toint\Ph^{(1)}_I \Ph_J\Bigr)\Ph_K\tP^L
\Bigr\rangle +
\Bigl\langle\Ph_J\Bigl(\toint\Ph^{(1)}_I \Ph_K\Bigr)\tP^L
\Bigr\rangle +\Bigl\langle \Ph_J\Ph_K
\Bigl(\toint\Ph^{(1)}_I \tP^L\Bigr)\Bigr\rangle =0.
\ee
Combined with (\ref{3pt}) and (\ref{rep}) this indeed gives (\ref{jacobi}).
Notice that it is irrelevant which operator actually played the role of the
current.

In a similar way one can study more general Ward-identities with more
operator insertions. This was done recently in refs.\ \cite{IK} and \cite{WZ},
where it was noted that  the symmetries can act non-linearly on the
vertex operators.
The origin of this fact is that the currents $\Ph^{(1)}_I$
are conserved only up to a BRST-derivatives, which after `partial integration'
produces higher order contact terms. Thus, a Ward-identity is actually a
statement about the decoupling of a total BRST-derivative in which
none of the operators plays a special role; this is the point of view
we will take in the next section.

\newsection{Ward identities}

String amplitudes are usually written as integrals
of correlation functions of two-forms $\Ph^{(2)}$ on a genus $g$
surface $\Sigma$. A more symmetric and for our purpose more convenient
way to represent the amplitudes is to use only zero-forms and extra
$b$-field insertions. In this way we treat the integrations over the
positions of the operators on equal footing with the other moduli of surface
$\Sigma$. Thus we write the $g$-loop $(r\!+\!s)$-point amplitude
as follows
\be
\label{ampl}
\Bigl\langle \Ph_{I_1}\ldots \Ph_{I_r}\tP^{J_1}\ldots\tP^{J_s}\Bigr\rangle_g=
\int_{{\cal M}_{g,r+s}} \biggl\langle\!\!\!\biggl\langle
\prod_{i\in {\cal I}}\Ph_{I_i}\prod_{j\in {\cal J}}\tP^{J_j}
\prod_{\alpha=1}^{3g-3+r+s} (\mu_\alpha,b)
(\mubar_\alpha,\bbar)
\biggr\rangle\!\!\!\biggr\rangle,
\ee
where ${\cal I}=\{1,\ldots,r\}$ and ${\cal J}=\{1,\ldots,s\}$.
The integral is over the moduli space ${\cal M}_{g,r+s}$
of genus $g$ surfaces with $r\!+\!s$ (marked) punctures,
and $\mu_\alpha$ represent the Beltrami-differentials, which are integrated
over the surface together with the $b$-fields.
The double brackets on the r.h.s. indicate the correlation function in the
2d field theory of the ghost and matter fields.
By ghost counting  the combined ghost-number of the states and
anti-states must add up to $2r\!+\!2s$ to get a non-zero result.

The requirement that unphysical modes in string theory
decouple is usually stated as the condition that amplitudes
containing BRST-trivial states vanish, \ie
\be
\label{Ward}
\Bigl\langle [Q,\Lambda]_{\pm} \prod_{i}\Ph_{I_i}\prod_{j}\tP^{J_j}\Bigr\rangle
= 0.
\ee
Here the amplitude is defined as in (\ref{ampl}).
In the following we assume that (\ref{Ward}) holds for all dimension zero
operators $\Lambda$ that satisfy $\bm|\Lambda\rangle=0$.

The naive reasoning for why (\ref{Ward}) is true
is as follows. One  `partially integrates' the
BRST-derivative and because $Q$ commutes with all $\Ph_I$ and $\tP^J$
it only acts on the $b$-insertions.
This produces an insertion of the stress-tensor $T=[Q,b]_+$.
Thus by a standard argument, the result is a total derivative
on moduli space and therefore naively vanishes.
In this naive argument we ignored the possible contributions that
come from the boundary components of moduli space. It is a non-trivial
fact that these contributions cancel, and this is what leads to the
Ward-identities. We will analyse the above
manipulations for a particularly convenient choice for
$\Lambda$ namely $\Lambda=1$. For this choice
(\ref{Ward}) is clearly true because $[Q,1]=0$, and secondly,
as we will see the different contributions at the boundary only
involve physical states.

First we consider the boundary components of ${\cal M}_{g,r+s}$
where the surface splits in to two surfaces
$\Sigma_1$ and $\Sigma_2$ with genera  $0\leq g_1,g_2\leq g$, $g_1+g_2=g$,
and the operators are distributed over $\Sigma_1$ and $\Sigma_2$
so that $r_1+r_2=r$ and $s_1+s_2=s$; in fig.\ 1 we have drawn this
situation for a one-loop $4$-point amplitude. In going to the boundary of
${\cal M}_{g,r+s}$ we take the length of one of the
cycles to zero and consequently only states with scaling dimension
zero give a finite contribution in the factorization expansion.
The remaining moduli can be recognized as the moduli of $\Sigma_1$ and
$\Sigma_2$, except for one `angular' parameter being the `twist' associated
with the pinched cycle. The integration over this modulus, together
with the corresponding $b$-ghost insertion, yields the BRST-invariant
operator
\be
\Pi=(b_0-\bbar_0)\delta(L_0-\Lbar_0).
\ee
There exists a basis of the Hilbert space $\cal H$ so that the
basis elements that are not annihilated by $Q$ are orthogonal to
the physical states and conjugate to the BRST-trivial
states. We can use this fact to write
the operator $\Pi$ as
\be
\label{factor}
\Pi=
\sum_K|\tP^K\rangle\langle\Ph_K| +\sum_K|\Ph_K\rangle\langle\tP^K|
\ + \quad \mbox{a $Q$-commutator.}
\ee
where the sum is only over the physical states.
The $Q$-commutator term in (\ref{factor})
can be dropped because of the assumption that
(\ref{Ward}) holds for all proper zero-forms  $\Lambda$. We
conclude therefore that only physical states give a contribution
to the Ward-identity.

For genus zero the boundary components of ${\cal M}_{0,r+s}$
are all of the type described above and are labeled by the
way the operators are distributed over the two surfaces $\Sigma_1$
and $\Sigma_2$.
Summing the different contributions leads to the following tree-level
Ward identities\footnote{After we finished our calculations
we noticed that this relation for the special case $s=1$  also
appeared in \cite{WZ}, where a similar derivation was given and
it was noted that it defines a so-called homotopy lie-algebra.}
\be
\label{WI}
\sum_K\sum_{{{\cal I}_1 \sqcup {\cal I}_2 ={\cal I}\atop {\cal J}_1 \sqcup
{\cal J}_2 ={\cal J}}}\Bigl\langle
\prod_{i\in {\cal I}_1} \Ph_{I_i}\prod_{j\in {\cal J}_1} \tP^{J_j}
\,\Ph_K\Bigr\rangle_{\strut{0}}
\Bigl\langle \,\tP^K\prod_{i^\prime\in {\cal I}_2} \Ph_{I_{i^\prime}}
\prod_{j^\prime\in {\cal J}_2} \tP^{J_{j^\prime}}\Bigr\rangle_{\strut{0}}=0,
\ee
where the sum is over all subdivisions ${\cal I}_1, {\cal I}_2$ and
${\cal J}_1, {\cal J}_2$ of the sets ${\cal I}$ and ${\cal J}$.
As a special case this contains the Jacobi identity (\ref{jacobi})
and the relation (\ref{relat}). Another interesting case is when we have
one `current' $J_a$ and for the rest  operators with $\gl=2$.
When the symmetry is linearly realized, only the three-point
amplitudes of the current $J_a$ are non-vanishing.
It is easily seen that in this case (\ref{WI}) reduces to the more
familiar form of a linear Ward-identity.

For higher genus surfaces there is one other type of
boundary component ${\cal M}_{g,r+s}$, namely when one of the
handles is pinched (see fig.\ 2). Following a similar argument
as sketched above we find that in this case the contribution to
the Ward-identity can be expressed as a (graded) trace
of the operator $\Pi$ times a BRST invariant operator that represents the rest
of the surface. Again by using (\ref{factor}) we can,
at least formally, reduce this trace to a sum over physical states.
This leads to the following extension of (\ref{WI}) to loop
amplitudes
\nopagebreak
\begin{eqnarray}
\label{WI2}
0\is
\sum_K \Bigl\langle \Ph_K\,\tP^K\,\prod_{i\in {\cal I}}
\Ph_{I_i}\prod_{j\in{\cal J}}\tP^{J_j}\Bigr\rangle_
{\strut{g-1}} +
\\[3mm] \nonumber
& &
+
\sum_{K\atop g_1+g_2=g}
\sum_{{{\cal I}_1 \sqcup {\cal I}_2 ={\cal I}\atop {\cal J}_1 \sqcup
{\cal J}_2={\cal J}}}\Bigl\langle \prod_{i\in {\cal I}_1}
\Ph_{I_i}\prod_{j\in {\cal J}_1} \tP^{J_j}
\Ph_K\Bigr\rangle_{\strut{g_1}}
\Bigl\langle \tP^K\,\prod_{i^\prime\in {\cal I}_2} \Ph_{I_{i^\prime}}
\prod_{j^\prime\in {\cal J}_2} \tP^{J_{j^\prime}}\Bigr\rangle_{\strut{g_2}},
\end{eqnarray}
where the first term comes from the degeneration of a handle.

How much can we learn from these Ward-identities?  Are they sufficient,
for example, to compute all string amplitude starting from, say, the
genus zero three and four-point functions? The answer to this question of
course depends on how large the symmetry group is, and whether it acts
non-trivially on all the physical operators. For the uncompactified
flat background the recent results of Klebanov \cite{IK},
who reproduced all the tachyon amplitudes using Ward-identities,
are an indication that this might be true. However, as will be
explained in the next section, the Ward-identities
(\ref{WI}) and (\ref{WI2}) have a large gauge-invariance,
and therefore to get a unique solution one in general needs
more information.

An interesting special case of (\ref{WI2}) is
\be
\label{tadpole}
f_{IJ}{}^K\langle \Ph_K\rangle_1 =-\sum_K\langle
\Ph_I\Ph_J \Ph_K\tP^K\rangle_0.
\ee
which shows that the one-loop tadpoles do not have to be invariant under
the symmetry when the gauge-group acts non-linearly at
tree-level. It also indicates that the classical symmetries are
modified by the quantum corrections.

\newsection{The Master Equation}

In this section we will recast the Ward-identities in a more
compact form as constraints on the partition function.
 Let us
introduce coupling constants $\rg^I$ and $\tg_I$ for the states and
anti-states and consider the perturbed action
\be
\label{pertu}
S^\prime=S+\sum_I \rg^I\!\int\Ph_I
+\sum_I \tg_I\!\int \tP^I.
\ee
Depending on whether the ghost number of $\Ph_I$
is even or odd the couplings $\rg^I$ are commuting or anti-commuting
variables. The couplings $\tg_I$ have the
opposite `statistics' from their anti-partners $\rg^I$.
More explicitly,
\newcommand{\ga}{n_{ag}}
\ba
g^I \is s^\alpha,\, c^a,\, t^i,\,  d^A\nonu
\tg_I \is \widetilde{s}_\alpha,\, \widetilde{c}_a,\, \widetilde{t}_i,
\, \widetilde{d}_A
\ea
represent the coupling for the states (\ref{fields}) and
anti-states (\ref{antifields}) respectively.
Here $s^\alpha, t^i,\widetilde{c}_a$, and
$\widetilde{d}_A$ are even and $c^a, d^A, \widetilde{s}_\alpha$ and
$\widetilde{t}_i$ are odd variables.
We note the couplings $c^a$ are in one to one correspondence with the
(gauge-)symmetries generated by $J_a$ and can be identified with the on-shell
modes of the space-time ghosts.
The other couplings have similar space-time
interpretations, for example $s^\alpha$ are the ghost for ghosts, etc.
In fact, one can define a space-time anti-ghost number $\ga$ by
\ba
\ga({g^I})\is\gl(\Ph_I)-2, \nonu
\ga({\tg_I})\is\gl(\tP^I)-2,
\ea
so that the couplings $t^i$ and $\widetilde{d}_A$ have $\ga=0$.

Our aim is to write the collection of all Ward-identities (\ref{WI})
and (\ref{WI2}) as a differential equation for the free energy $F$ as a
function of the couplings $\rg^I$ and $\tg_I$.
The free energy $F$ is the generating function for all connected
string amplitudes
\be
\label{Fdef}
\Bigl\langle \Ph_{I_1}\ldots \Ph_{I_r}\tP^{J_1}\ldots\tP^{J_s}\Bigr\rangle=
-{\partial^{r+s}F(\rg^I,\tg_J)
\over \partial \rg^{I_1}\ldots\partial\tg_{J_s}}{}_{\strut{
|_{g^I=\ttg_J=0}}}
\ee
and has a genus-expansion
\be
F=\sum_g \lambda^{2g} F_g
\label{expans}
\ee
in terms of the string coupling constant $\lambda$.
It easily seen that $F$ is an even function and has total antighost number
$\ga=0$. In this section $\langle\cdots\rangle$ denotes the full
connected amplitude including the loop contributions, unless otherwise
indicated.

\newsubsection{The classical master action}

When we turn on the couplings ghost-number is no longer conserved, and
the operators $\Ph_I$ and $\tP^J$ can
acquire non-vanishing one-point functions on the sphere.
The tree-level Ward-identities (\ref{WI}) can be integrated several
times to obtain an expression in terms of these one-point functions
\be
\label{sum}
\sum_K \langle\Ph_K\rangle_0\langle\tP^K\rangle_0=0,
\ee
which holds for arbitrary values of couplings. In fact, this single equation
contains the same information as the infinite set of identities (\ref{WI}),
which simply follow by differentiating (\ref{sum}) $r$ times
with respect to the couplings $\rg^I$, and $s$ times with respect to $\tg_J$.
The `statistics' of the couplings is important to get the correct signs
for the different terms.

To write (\ref{sum}) in terms of the tree level free energy $F_0$
we introduce the so-called anti-bracket. It is defined by
\be
\{A,B\} = \sum_I \Bigl({\partial A\over\partial \tg_I}{\partial B\over
\partial \rg^I}\pm {\partial B\over\partial \tg_I}{\partial A\over
\partial \rg^I}\Bigr),
\ee
where the plus-sign occurs only when both $A$ and $B$ are commuting.
Equivalently we may define the anti-bracket through
\be
\{\tg_J,\rg^I\}=\delta_J{}^I,\qquad\qquad \{\rg^I,\rg^J\}=\{\tg_I,\tg_J\}=0,
\ee
supplemented with the rule that the anti-bracket acts as a
derivation.
So in the anti-bracket the fields $\rg^I$ are conjugate to the
anti-fields $\tg_I$, but in contrast with the Poisson-bracket
conjugate variables have opposite `statistics'.

Using the anti-bracket the relation (\ref{sum}) can be
compactly written as
\be
\label{BV0}
\{F_0,F_0\}=0,
\ee
which summarizes all tree level Ward identities.
Equation (\ref{BV0}) is familiar from the Batalin-Vilkovisky formalism,
where it is known as the classical master equation.
This is no coincidence: the BV-formalism is developed precisely
to give a general framework for writing Ward-identities in arbitrary
gauge theories. The free energy $F_0$ plays the role of the `classical'
master action for all fields and anti-fields and at the same time it is the
generator of BRST-transformations. Thus, equation (\ref{BV0})
expresses the invariance of the `action' $F_0$ under the
BRST-transformations
\ba
\delta \rg^I &= \epsilon \{F_0,\rg^I\} =&
\epsilon{\partial F_0\over\partial \tg_I},
\nonu
\delta \tg_J&= \epsilon \{F_0,\tg_J\}=&
\pm\epsilon {\partial F_0\over\partial \rg^J},
\ea
as well as the nilpotency of the
BRST charge. To make this interpretation a bit more apparent we note
that, in lowest order, $F_0$  is given by
\be
\label{F}
F_0=c^a t^i f_{ai}{}^j\widetilde{t}_j + \half c^ac^b f_{ab}{}^c
\widetilde{c}_c+\dots-\half t^it^jC_{ij}{}^A\widetilde{d}_A+\ldots
\ee
In the first two terms we recognize the conventional BRST-charge:
the $c^a$ are the modes of the space-time
ghost fields and that the coefficients $f_{ai}{}^j$ and $f_{ab}{}^c$ are the
generators and structure constants of the physical symmetries.
For symmetries that are linearly realized are the
only terms involving the ghosts $c^a$, but in general there will be
more terms.

Due to the analogy with the BV-formalism it is natural to think of the
tree-level free energy $F_0$ as being (part of) the classical space-time
action of the 2d string. Indeed, its equation of motion
\be
\label{beta}
{\partial F_0\over \partial \widetilde{d}_A}= -\half
t^i t^j C_{ij}{}^A +\ldots =0
\ee
coincides in first order with the quadratic $\beta$-function condition,
and one may conjecture that the higher order terms
reproduce the full $\beta$-functions. This suggest that $F_0$ is
closely related to Zamolodchikov's $c$-function.

\newsubsection{The quantum master equation}

The Ward-identities (\ref{WI2}) for the loop amplitudes
have a similar structure, but contain the contributions due
the pinching of a handle. In fact, this has a very natural
natural interpretation within the BV-formalism,
namely  it leads to what is called
the quantum master equation.
It involves the quadratic differential operator
\be
\Delta=\sum_I{\partial^2\over\partial \rg^I\partial\tg_I}.
\ee
Notice that $\Delta$ is nilpotent,
{\ie} it satisfies $ \Delta^2=0.$
The quantum master equation reads in terms of complete
free energy $F$
\be
\label{master}
-\lambda^2\Delta F+\half\{F,F\}=0.
\ee
The second term describes the splitting of the surface in two component
while the first term originates from the pinched handle. This becomes more
evident if we rewrite it again in terms of the amplitudes.
The master equation then becomes
\be
\label{mast}
\lambda^2 \sum_I  \langle \tP_I\Ph^I\rangle+\sum_I
\langle\tP_I\rangle \langle \Ph^I\rangle =0,
\ee
which again is valid for arbitrary values of the couplings.
The first term in (\ref{mast}) can be non-zero even when we put all
couplings to zero. In this case also the one-point functions must
be non-zero, and consequently the loop corrections to the
`$\beta$-function' (\ref{beta})  will modify the classical solutions.
This phenomenon is known as the `Fischler-Susskind mechanism' \cite{FS}.

We can even further simplify the master equation
by expressing it in the full partition function
\be
Z(\rg^I,\tg_I) =e^{-{1\over\lambda^{2}}F(g^I,\ttg_I)}.
\ee
It then takes the form
\be
\label{DZ}
\Delta Z=0,
\ee
which looks misleadingly simple but as is clear from our derivation,
it contains all the information about the Ward-identities (\ref{WI2}).
The fact that $\Delta^2=0$ is a direct consequence of the nilpotency
of the BRST-charge $Q$ on the world sheet.
We observe that (\ref{DZ}) is invariant under
\be
\label{gauge}
Z\rightarrow Z+\Delta Y,
\ee
where $Y$ an arbitrary odd function of the couplings with $\ga=1$. This
gauge-invariance is related to the fact that we can change the physical
fields $\Ph_I$ by total BRST-derivatives of `bad' operators, which do
not decouple. At tree level this redundancy corresponds
to reparametrizations in the space of couplings $(\rg^I,\tg_J)$
that preserve the anti-bracket.
Therefore we believe that two partition functions
that are related by a gauge transformation of the form
(\ref{gauge}) must be considered to be equivalent.

Although naively the master equation (\ref{DZ}) does not depend on the
background we emphasize that its derivation used a particular fixed
background. The partition function is only formally defined as a
perturbative expansion in the couplings around $\rg^I=\tg_I=0$. In this point
the first and second derivatives of $F$ vanish, but of course this
is not true for other values of the couplings. This makes clear that if we
compute the partition function perturbatively for some other background
it will not produce the same answer. It is possible that the answers are
the same after a gauge transformation (\ref{gauge}),
but we do not know whether this is true in general.

Finally, we note that equation (\ref{DZ}) is quite reminiscent of
the Virasoro  and $W$-constraints for $c\leq 1$ gravity \cite{DVV1}, and
also its derivation is analogous to the way the Virasoro constraints
were found in topological gravity \cite{VV}. An important difference is
that the master equation gives only one constraint, and is in itself
not sufficient to fix $Z$, even up to equivalence.
It seems natural to expect that the master equation also holds
non-perturbatively; in this case it may even tell us something
about the nature of the non-perturbative effects \cite{SS}.

\newsection{Concluding remarks}

In this paper we have described a general framework for studying the
symmetries of 2D string theory. We realize that our discussion has been
rather formal, for example, we implicitly assumed that all amplitudes
and all the sums over states are well defined and finite.
In practice  all kinds of divergences can arrise,
but we believe that the master equation itself can help in dealing
with these divergences. It is useful, therefore, to apply and
verify these ideas more explicitly in concrete situations.

An important open problem is how to relate different backgrounds.
The partition function $Z$ was defined through its perturbative
expansion in the coupling constants for the physical states.
The main obstacle in defining $Z$ as a function
of the background is the fact that the spectrum of physical
states varies rather discontinuously.
This brings us back to our discussion of section 2.3
where we studied the perturbations of the BRST-charge.
We found that the perturbed BRST-charge
can be expressed in terms of the symmetry charges as
\be
Q^\prime=Q+\sum_I g^I Q_I+\ldots.
\ee
The non-linear symmetry that is generated by these charges $Q_I$
is encoded in the master equation. This makes clear that by studying
the master equation we should also learn something
about the perturbations of the physical spectrum.

To make progress in this direction one could try to
enlarge the space of coupling constants, by including states
that become physical at other values of the couplings.
In this way one might be forced to go to the full closed string
field theory, but hopefully there is a more tractable intermediate
formulation. The idea that such a intermediate formulation
may exists has also been put forward in \cite{WZ}.
In this respect it is interesting to note that a
BV-equation already exists for
closed string field theory \cite{BZ}.
The master equation that we derived in this paper presumably corresponds
to the on-shell truncation of the full master equation of string field
theory.

We have put our discussion in the context of 2D string theory, because
at present it is the only non-trivial example of a string theory
with a large symmetry algebra. One can try to apply the same techniques to
$c\leq1$ models, or maybe even to $c\geq 1$. Unfortunately, the representation
theory of the Virasoro algebra with $c\geq 1$ indicates that it is not
possible to have many discrete states in higher dimensional models.
We note, however, that conformal invariance played only a minor role in
our derivations. It is quite possible that more general backgrounds exist
that are not necessarily based on conformal field theories. Such models can
have large symmetry algebras, and then our results would still
be applicable.

\begin{flushleft}
{\sc Acknowledgements}: I thank R. Dijkgraaf, I. Klebanov, H. Verlinde,
E. Witten, and B. Zwiebach for stimulating discussions and helpful
comments.
\end{flushleft}

{\renewcommand{\Large}{\large}

}

\newpage

\thispagestyle{empty}

{\sc FIGURE CAPTIONS:}
\bigskip

\begin{flushleft}
{\it Figure} $1$: A torus with $4$ marked
punctures that degenerates by pinching off a sphere
with $2$ punctures.
\end{flushleft}

\bigskip

\begin{flushleft}
{\it Figure} $2$: A torus with $4$ marked punctures that
degenerates by pinching a handle.
\end{flushleft}
\end{document}